
\input harvmac

\def\rappendix#1#2{\global\meqno=1\global\subsecno=0\xdef\secsym{\hbox{#1.}}
\bigbreak\bigskip\noindent{\bf Appendix. #2}\message{(#1. #2)}
\writetoca{Appendix {#1.} {#2}}\par\nobreak\medskip\nobreak}

\def\ack{\bigbreak\bigskip\bigskip\centerline{{\bf Acknowledgements}}\nobreak}


\def\MPL#1#2#3{{Mod.~Phys.~Lett.} {\bf A#1} (19#2) #3}

\def\PL#1#2#3{{Phys.~Lett.} {\bf B#1} (19#2) #3}
\def\NP#1#2#3{{Nucl.~Phys.} {\bf B#1} (19#2) #3}
\def\CMP#1#2#3{{Commun.~Math.~Phys.} {\bf #1} (19#2) #3}
\def\LMP#1#2#3{{Lett.~Math.~Phys.} {\bf #1} (19#2) #3}
\def\TMP#1#2#3{{Theor.~Math.~Phys.} {\bf #1} (19#2) #3}
\def\TMF#1#2#3{{Teor.~Mat.~Fiz.} {\bf #1} (19#2) #3}


\def\frac#1#2{{#1 \over #2}}
\def\inv#1{{1 \over #1}}
\def\ha{{1 \over 2}}
\def\der{\partial}
\def\vev#1{\langle #1 \rangle}
\def\ket#1{ | #1 \rangle}
\def\->{\rightarrow}     \def\<-{\leftarrow}
\def\<{\langle}          \def\>{\rangle}
\def\[{\left [}          \def\]{\right ]}
\def\({\left (}          \def\){\right )}
\def\abs#1{\left\vert #1 \right\vert}
\def\bun#1/#2{\leavevmode
   \kern.1em \raise .5ex \hbox{\the\scriptfont0 #1}%
   \kern-.1em $/$%
   \kern-.15em \lower .25ex \hbox{\the\scriptfont0 #2}%
}

\def\apar{\noalign{\vskip 2mm}}


\def\op#1{\mathop{\rm #1}\nolimits}
\def\dim{\op{dim}}
\def\min{\op{min}}


\def\bg{{\bf g}}

\def\bZ{{\bf Z}}

\def\a{\alpha}
\def\b{\beta}
\def\d{\delta}		\def\D{\Delta}

		\def\G{\Gamma}
\def\la{\lambda}	        \def\La{\Lambda}
\def\o{\omega}		
\def\p{\phi}            \def\P{\Phi}
\def\ps{\psi}           
\def\r{\rho}
	
\def\t{\theta}		
\def\vp{\varphi}


\def\cw{conformal weight}
\def\minimal{$N=2$ minimal model}
\def\eb{{\bar\eta}}
\def\fg{\CU_\eb}
\hyphenation{functions}


\lref\rEY{T.~Eguchi and S.-K.~Yang, \MPL{4}{90}{1653}. }
\lref\rW{E.~Witten, \CMP{118}{88}{411}; \NP{340}{90}{281}.}
\lref\rDVV{R.~Dijkgraaf, E.~Verlinde and H.~Verlinde, \NP{352}{91}{59};
           PUPT-1217, IASSNS-HEP-90-80, (November, 1990).}
\lref\rK{T.~Kawai, \PL{259}{91}{460}; {\bf B261} (1991) 520E.}
\lref\rEKMY{T.~Eguchi, T.~Kawai, S.~Mizoguchi and S.-K.~Yang,
            KEK-TH-303, KEK preprint 91-111 (September, 1991).}
\lref\rNY{M.~Ninomiya and K.~Yamagishi, \PL{183}{87}{323}.}
\lref\rG{D.~Gepner, \NP{290}{87}{10}.}
\lref\rSS{A.~Schwimmer and N.~Seiberg, \PL{184}{87}{191}.}
\lref\rLVW{W.~Lerche, C.~Vafa and N.P.~Warner, \NP{324}{89}{427}.}
\lref\rEHY{T.~Eguchi, S.~Hosono and S.-K.~Yang, \CMP{140}{91}{159}.}
\lref\rKUY{T.~Kawai, T.~Uchino and S.-K.~Yang, in preparation.}
\lref\rBMP{P.~Bouwknegt, J.~McCarthy and K.~Pilch, \CMP{131}{90}{125}.}
\lref\rZF{A.B.~Zamolodchikov and V.A.~ Fateev, \TMP{71}{88}{451}
           \hfil\break
          [\TMF{71}{87}{163}].}
\lref\rGS{P.~Goddard and A.~Schwimmer, \PL{206}{88}{62}.}
\lref\rGii{D.~Gepner, \CMP{141}{91}{381}.}
\lref\rWii{E.~Witten,  IASSNS-HEP-91-25 (June, 1991).}
\lref\rI{K.~Intriligator, HUTP-91/A041 (August, 1991).}
\lref\rNW{D.~Nemeschansky and N.P.~Warner, USC-91/031 (October, 1991).}
\lref\rD{E.~Date, M.~Jimbo, A.~Kuniba, T.~Miwa and M.~Okado, \LMP{17}{89}{69}.}
\lref\rS{H.~Saleur,  YCTP-p38-91 and YCTP-p39-91 (October, 1991).}
\lref\rV{See for a review, C.~Vafa, HUTP-91/A059 (November, 1991).}


\Title{\vbox{\baselineskip12pt\hbox{KEK-TH-311}
                             \hbox{KEK preprint 91-190}
                              \hbox{December 1991}}}
{\vbox{\centerline{Higher-Rank Supersymmetry}
\bigskip
\centerline{and}
\bigskip
        \centerline{Topological  Field Theory*}}}
       \footnote{}{*To appear in the proceedings of YITP Workshop on
                ``Developments in Strings and Field Theories'' held at Kyoto,
           Japan, on September 9-12, 1991}

\centerline{Toshiya Kawai, Taku Uchino and Sung-Kil Yang}
\it
\vskip.5in\centerline{National Laboratory for High Energy Physics (KEK)}
\vskip.3in\centerline{Tsukuba, Ibaraki 305, Japan}
\rm
\vskip .6in
\centerline{\bf Abstract}
\vskip.3in
The $N=2$ minimal superconformal model can be twisted yielding an example of
topological conformal field theory.
In this article we investigate  a Lie theoretic extension  of this process.

\Date{}

\newsec{Introduction}

The twisting of the $N=2$ minimal superconformal model \rEY\ gives rise to
a fundamental class of topological field theories \rW.
Topological conformal field theories (TCFT) realized as the topological
version of certain $N=2$ models exhibit remarkable properties such as the
existence of the flat coordinate system and the prepotential \rDVV.
It is thus quite interesting to ask to what extent one can generalize the
relationship between TCFT and $N=2$ superconformal theories. In this
contribution we wish to point out that there is a natural Lie theoretic
extension of the $N=2$ superconformal algebra which makes us possible to
construct a wider class of TCFT.

This paper is organized as follows. In sect.2 we introduce the higher-rank
generalization of the \minimal. In sect.3 we show that the original idea of
the chirality in $N=2$ theory can be  extended in our model.
In sect.4, applying the twisting procedure, we discuss topological properties.
In sect.5 we sketch how to evaluate multiple integrations in order to
demonstrate the BRST exactness of the twisted stress
tensor. In sect.6 we present some concluding remarks.

\newsec{Higher-Rank Supersymmetry}

The close resemblance between the \minimal\ and the $SU(2)$ WZNW model
has been well recognized. If one considers  an arbitrary simple Lie algebra,
 motivated by the $SU(2)$ paradigm, the `higher-rank' analog of the \minimal\
appears \refs{\rK,\rEKMY}. This is our model based on which we shall construct
 TCFT.
The model is described by a familiar system
${(\bg{\rm -parafermion})_k\times({\rm boson})^n}$
with the special values of Coulomb gas parameters
\eqn\CGps{\a_+=\sqrt{\frac{k+g}{kg}},\quad \a_+\a_-=-\inv{k},
         \quad  \a_0=\a_++\a_-.}
Here $\bg$ is a simple Lie algebra of rank $n$,  the {\it level} $k$ is
 a positive integer, and $g$ is the dual Coxeter number of $\bg$.
(We follow the convention of \rEKMY.)
The \minimal\ corresponds to the case $\bg=\CA_1$.

The existence of   `exotic supersymmetry' in this system has been expected
 through the analysis of the branching relations associated with
 the coset model $\bg_k\oplus U(1)^n/U(1)^n$ \rK.
 The generators of this symmetry are given by\foot{
Throughout this paper we suppress normal orderings.}
\eqn\susygen{ G^\a(z):=\ps_\a(z)e^{i\a_+\a\cdot\vp(z)},
           \qquad \a\in\D}
where  $\D$ is the set of roots of $\bg$,  $\ps_\a(z)$ are the generating
parafermions, and $\vp(z)=(\vp_1(z),\ldots,\vp_n(z))$ are free bosons with
their two-point functions  $\vev{\vp_a(z)\vp_b(w)}
\allowbreak=\allowbreak-\d_{ab}\allowbreak\log(z-w)$.
The stress tensor consists of the parafermion piece $T_{\rm para}(z)$ and
the free boson piece
\eqn\stress{\eqalign{T(z)&=T_{\rm para}(z)-\ha(\der\vp(z))^2\cr
                         &=\sum_{n\in\bZ}z^{-n-2}L_n\, .\cr}}
The central charge $c$ is given by $c=\frac{kd}{k+g}$ with $d=\dim\bg$.
Since $\ps_\a(z)$ has a conformal weight $1-\frac{\a^2}{2k}$ \refs{\rNY,
\rG},
the \cw\ of $G^\a(z)$ is $1+\frac{\a^2}{2g}$.
There are also $U(1)^n$ currents which we normalize by
\eqn\uone{J^a(z)=\frac{i}{\a_+}\der\vp_a(z)=\sum_{n\in\bZ} z^{-n-1}J^a_n,\quad
 a=1,\ldots,n\, .}
The chiral algebra generated by $G^\a(z)$, $T(z)$ and $J^a(z)$ is closed and
associative by construction. We will refer to this symmetry as the
higher-rank supersymmetry.
The structure of this algebra is, in general,  complicated due to
nonlocality as well as nonlinearity,
but, as we will see below, the twisting procedure can be
studied without much difficulty since we do not need the whole
algebra\foot{See, however, Appendix where we find
 a connection to  the Zamolodchikov-Fateev spin-\bun{4}/{3} current algebra,
when $\bg=\CA_2$.}.

We now describe the reason why we think of this chiral algebra as the
higher-rank extension of the $N=2$ algebra. The crucial ingredient in the
$N=2$ model is the spectral flow \rSS\ and the chiral ring \rLVW. These
properties
are in fact realized in our present model. Let us first consider the spectral
flow. The existence of $U(1)^n$ currents implies an $n$-parameter spectral
flow. We denote the flow operator by $\CU_\eb$ with $\eb$ being an
$n$-dimensional flow vector. One finds an inner automorphism of the algebra
\eqn\sflowi{\eqalign{
      \fg\, L_n \,\fg^{-1}&=L_n+\eb\cdot J_n+
         \frac{\abs{\eb}^2}{2\a_+^2}\,\d_{n,0}\, ,\cr
\apar
        \fg\, J_n^a \,\fg^{-1}&=J_n^a+\frac{\eb^a}{\a_+^2}\,\d_{n,0}\, ,\cr
\apar
        \fg\, G^\a_r \,\fg^{-1}&=G^\a_{r+\eb\cdot\a}\, ,\cr}}
where
\eqn\susymode{G^\a(z)=\sum_r z^{-r-\(1+\frac{\a^2}{2g}\)}\,G_r^\a\,.}
Specializing $\eb=\frac{\eta}{g}\r$ with $\r$ being half the sum of the
positive roots of $\bg$ yields a one-parameter flow
\eqn\sflowii{\eqalign{
 \fg\, L_n\, \fg^{-1}&=L_n+\frac{\eta}{g}\r\cdot J_n+
                  \frac{c}{24}\eta^2\,\d_{n,0}\, ,\cr
\apar
        \fg\, J_n^a\, \fg^{-1}&=J_n^a+\frac{c}{d}\eta\r^a\,\d_{n,0}\, ,\cr
\apar
        \fg\, G^\a_r\, \fg^{-1}&=G^\a_{r+\frac{\r\cdot\a}{g}\eta}\, ,\cr}}
where the strange formula $12 \r^2=gd$ has been used. As in $N=2$
theories this flow is important when we consider the analogs of the chiral
ring and the Ramond ground states of the present system in the next section.

\newsec{Chiral Primary Fields}

In $N=2$ superconformal field theory chiral primary fields play a
distinguished role \rLVW.
In particular, they become BRST invariant physical observables after
twisting \refs{\rEY{--}\rDVV}.
To find analogs of chiral primary fields, for which we use the same
terminology in the following, let us define the NS sector as the set of
 fields $A_{\rm NS}(z)$ which are local with respect to the fractional-spin
currents
$G^\a(z)$. Then
\eqn\NSmode{G^\a(z)A_{\rm NS}(0)=\sum_{r\in \bZ-\frac{\a^2}{2g}}
            z^{-r-\(1+\frac{\a^2}{2g}\)}\, \(G_r^\a A_{\rm NS}\)(0)\,.}
The NS primary states are annihilated by all the positive modes, {\it i.e.}
 $L_n$, $J_n^a$ and $G_{n-\frac{\a^2}{2g}}^\a$ with $n>0$.
Chiral fields  $\P$ are those in the NS sector obeying\foot{
We note that there are as many similar sets of conditions as the order of
the Weyl group of $\bg$ corresponding to different choices of the simple root
 system. Hence for each such choice one can repeat the whole story in the
sequel.}
\eqn\chiralcond{\oint_0 dz\,G^{\a_i}(z)\P(0)=0\, ,\qquad i=1,\ldots,n\, ,}
for each simple root $\a_i$ of $\bg$. Chiral primary fields satisfy both
\chiralcond\ and the primary condition.  These are explicitly obtained as
\eqn\cfield{\P_\La (z)=\p^\La_\La(z)e^{-i\a_-\La\cdot\vp(z)},\qquad
\La\in P_+^k}
where $\p^\La_\La$ is a parafermionic primary field with a conformal weight
\eqn\pfpweight{\inv{2(k+g)}(\La+2\r)\cdot\La-\frac{\La^2}{2k}\, ,}
and $P_+^k$ is the set of dominant weights appearing in the level $k$
 WZNW model. Although one could consider $\P_\La$ for any values of the Coulomb
 gas parameters as a deformation of the highest weight primary field of the
 WZNW model, one special feature arising from \CGps\ is that the \cw\   and
$U(1)^n$ charges of $\P_\La$ are linearly related:
\eqn\chiralcw{L_0\ket{\P_\La}=\inv{g}\r\cdot J_0 \ket{\P_\La}\, ,}
with
\eqn\chiralcge{J_0^a\ket{\P_\La}=\frac{g\La^a}{k+g}\ket{\P_\La}\,.}
Then one can adopt  the conventional argument
(found {\it e.g.} in \rLVW) and deduce  that there are no short
 distance singularities between two chiral primary fields:
\eqn\cOPE{\P_\La(z)\P_{\La'}(w)\sim\P_{\La+\La'}(w)\, ,}
where we should understand that $\P_{\La+\La'}(w)=0$ if
$\La+\La'\not\in P_+^k$. Thus our chiral primary fields define a finite
nilpotent ring.

Let us apply the flow \sflowii\ with $\eta=1$ to the NS sector.
The chiral primary states $\ket{\P_\La}$ are then mapped onto the R ground
states which have a  \cw\ $h$ and $U(1)^n$ charges $q$ given by
\eqn\Rcwcge{h=\frac{c}{24},\qquad
            q= \frac{g}{k+g}(\La+\r)-\r\,.}
These states are indeed responsible for the non-vanishing contribution to
the `generalized index' \rK.

\newsec{Topological Version}

Starting with the higher-rank supersymmetric model introduced in the previous
sections let us now construct TCFT. The first step is to twist the model
through the redefinition of the stress tensor
\eqn\twist{T(z)\-> {\hat T}(z)=T(z)+\inv{g}\r\cdot J(z)\,.}
The new stress tensor ${\hat T}(z)$ satisfies the Virasoro algebra with a
vanishing central charge.
Among the supercurrents $G^\a(z)$ with $\a\in\D$ we see that the currents
$G^{\a_j}(z)$ with simple roots $\a_j$ acquire conformal weights $1$ with
respect to
the new stress tensor. Hence BRST operators  in the topological version
will be constructed from $G^{\a_j}(z)$.
The BRST structure of the theory is governed by a directed graph
essentially determined by the affine Weyl group. One associates a BRST
charge to each arrow, the explicit form of which is given by a multiple
integral of $G^{\a_j}$  and depends on the location of the arrow in the graph.

In TCFT the stress tensor is expressed as a BRST commutator
\eqn\tstress{{\hat T}(z)=\{Q,G(z)\}\,.}
For $G(z)$ we take the current $G^{-\t}(z)$ whose \cw\ is $2$ with respect
to ${\hat T}(z)$ and the explicit expression of the BRST commutator
was proposed in \rEHY:
\eqn\screen{\eqalign{
&{\hat T}(z)=({\rm const.})\,T_s(z)\cr
\apar
&T_s(z):=\int\cdots\int_\G\, \prod_{i=1}^n\prod_{j=1}^{a_i} dz_i^{(j)}\,
G^{\a_i}(z_i^{(j)})G^{-\t}(z),}}
where $\t=\sum_{i=1}^n a_i\a_i$ is the highest root of $\bg$.
The contour $\G$ of the integrals must be chosen appropriately \rBMP.
The evaluation of $T_s(z)$ will be studied in the next section.

The physical states in our topological theory are realized as the non-vanishing
 BRST cohomology. Though the analysis of the BRST complex is quite complicated,
the index calculation based on the branching relation shows that the chiral
primary fields $\P_\La(z)$, $\La\in P_+^k$ turn out to be the basic physical
observables. In fact the BRST invariance of $\P_\La(z)$ immediately follows
from the chirality condition \chiralcond\  since the action of BRST charges
on $\P_\La(z)$ is given by
\eqn\qchiral{0=\{Q_{(j)},\P_\La(z)\}=\oint_zdw\, G^{\a_j}(w)\P_\La(z),
\quad j=1,\ldots,n.}
All these states have zero conformal
weights with respect to ${\hat T}(z)$ by virtue of \chiralcw.
Thus the physical states are labeled by the $U(1)^n$ charge \chiralcge.

Let us take a look at correlation functions. Since the stress tensor takes the
BRST exact form \screen, correlators of the basic physical operators are
independent of their world sheet positions, and hence there is no notion of
distance in the theory.  This is a familiar phenomenon observed in
twisted $N=2$ theories \refs{\rEY{--}\rDVV}.
We also note that upon twisting we have coupled the system to a `background
charge at infinity'.  The selection rule of $U(1)^n$ charges  arising
from this features our topological field theory.  We intend to further
discuss the properties of correlation functions, including perturbed behaviors,
in a subsequent paper \rKUY.

\newsec{Multiple Integrals}

In this section we briefly comment on the validity of \screen.
For simplicity, consider the case $\bg=\CA_n$ and $k=1$. By a standard
calculation we obtain
\eqn\Atype{\eqalign{
          T_s(z)=\int_\G\,\prod_{j=1}^ndz_j\,
           &\[\prod_{j=1}^{n-1}(z_j-z_{j+1})(z_1-z)(z_n-z)\]^{-\a_+^2}\cr
\apar
&\times\exp\[i\a_+\sum_{j=1}^n\a_j\cdot(\vp(z_j)-\vp(z))\]\,,}}
where $\a_+^2=\frac{n+2}{n+1}$ and we have used the fact
 $\t=\sum_{j=1}^n\a_j$. If we make a change of variables
$z_j\->z+\prod_{l=1}^ju_l$, $j=1,\ldots,n$ and integrate over $u_1$, we find,
after standard manipulations of contours, that
\eqn\Atypeii{\eqalign{
            T_s(z)\propto\prod_{j=2}^n\,&\[\oint du_j\,
              u_j^{\(\frac{j}{n+1}-1\)-1}(1-u_j)^{-\inv{n+1}-1}\]\cr
\apar
      &\times\Bigg[-\ha\Bigg(\a_1\cdot\der\vp(z)+\sum_{j=2}^n(u_2\cdots u_j)
           \a_j\cdot\der\vp(z)\Bigg)^2\cr
\apar
       &+\frac{i}{2}(n+1)\a_0\Bigg(\a_1\cdot\der^2\vp(z)+
              \sum_{j=2}^n (u_2\cdots u_j)^2\, \a_j\cdot\der^2\vp(z)
               \Bigg)\Bigg]\,.\cr}}
By repeatedly applying the recursion property of the beta function,
\eqn\betafun{B(a,b)=\oint dt\,t^{a-1}(1-t)^{b-1},
             \qquad B(a+1,b)=\frac{a}{a+b}B(a,b)\,,}
it is easy to see that
\eqn\Aemt{ T_s(z)\propto -\ha\sum_{i,j=1}^n(\o_i\cdot\o_j)
            (\a_i\cdot\der\vp)(\a_j\cdot\der\vp)(z)+
             i\a_0\r\cdot\der^2\vp(z)\,,}
where $\o_1,\ldots,\o_n$ are the fundamental weights of $\CA_n$ and
\eqn\metric{\eqalign{&\o_i\cdot\o_j=\min(i,j)-\frac{ij}{n+1}\, ,\cr
            &2\r=\sum_{i=1}^ni(n+1-i)\a_i\, .}}
Finally using the formula
\eqn\ortho{\sum_{i,j=1}^n(\o_i\cdot\o_j)(\a_i)^a(\a_j)^b=\d_{ab}\, ,}
which follows from $\sum_{a=1}^n(\a_i)^a(\o_j)^a=\d_{ij}$ and
$\sum_{i=1}^n(\a_i)^a(\o_i)^b=\d_{ab}$ we arrive at
\eqn\Aemtf{T_s(z)\propto-\ha(\der\vp(z))^2+i\a_0\r\cdot\der^2\vp(z)\,.}

For other Lie algebras we encounter more complicated integrals.
For instance, in the case $\bg=\CD_n$ we have to deal with lower moments
with respect to a two-variable Selberg density
to derive the desired result. Further details of the calculation will be
reported elsewhere \rKUY.

\newsec{Concluding Remarks}

We have seen that our higher-rank supersymmetric model is a fairly
natural extension  of the \minimal\ and its topological version
possesses all the properties characteristic to TCFT.

Let us  point out some interesting issues which remain to be properly
understood.
One may notice that the modified stress tensor ${\hat T}(z)$ is that
describing the $\bg_k\oplus\bg_0/\bg_k\simeq\bg_k/\bg_k$ coset theory.
According to recent works \refs{\rGii,\rWii} it has been established
that correlation functions in topological $\bg_k/\bg_k$ model yield
the fusion algebra of the WZNW model $\bg_k$. Furthermore, this result
can be understood in terms of the deformed chiral ring of topological $N=2$
theory \refs{\rI,\rNW}. Thus it will be significant to clarify whether the
present chiral primary ring, after certain deformation, has any relevance to
the $\bg_k$ fusion algebra \rKUY.

It should also be mentioned
that the higher-rank supersymmetric models are realized in the critical limit
of solvable vertex-type lattice models \rD.  Quite recently,
Saleur \rS\ has
observed an interesting structure of solvable lattice models whose critical
behaviors are described by the \minimal.  It seems important to seek a similar
correspondence in our higher-rank setting.

Finally we note that the parallelism with  $N=2$ supersymmetry
 is yet to be completed. What is crucially missing in  higher-rank
supersymmetry is its possible connection to certain geometry if there is
any. It would be very exciting if one can find a geometric interpretation
which could be an analog of the deep relation between Ricci-flat K{\" a}hler
manifolds and $N=2$ superconformal field theories \rV.

\ack
We thank E.~Date, T.~Eguchi  and Y.~Yamada for helpful discussions.

\rappendix{A}{}

The Zamolodchikov-Fateev spin-\bun{4}/{3} algebra \rZF\ is generated by two
spin \bun{4}/{3} currents $\ps(z)$ and $\ps^*(z)$ which  satisfy the
operator product expansions
\eqn\ZFalgebra{\eqalign{
&\ps(z)\ps(w)\sim\frac{\la}{(z-w)^{4/3}}\[\ps^*(w)+\ha(z-w)\der\ps^*(w)
+\cdots\]\, ,\cr
\apar
&\ps^*(z)\ps^*(w)\sim\frac{\la}{(z-w)^{4/3}}\[\ps(w)+\ha(z-w)\der\ps(w)
+\cdots\]\, ,\cr
\apar
&\ps(z)\ps^*(w)\sim\inv{(z-w)^{8/3}}\[I+\frac{8}{3c}(z-w)^2\,T(w)+
           \cdots\]\, .\cr}}
The associativity gives a constraint \rZF
\eqn\clrel{9\la^2c=4(8-c)\, ,}
which can be parametrized as
\eqn\parametrization{
       c=2\(1-\frac{12}{v(v+4)}\),\qquad
       \la^2=\inv{3}\frac{(v+2)^2}{(v-2)(v+6)} \, . }
The minimal unitary series corresponds to $v=2,3,\ldots$ and is realized by
the GKO coset  $(\CA_1)_4\oplus(\CA_1)_{v-2}/(\CA_1)_{v+2}$.

We present here  another realization of this algebra.
Consider $G^\a(z)$, $\a\in\D$, defined in \susygen\  for  $\bg=\CA_2$.
Note that all these currents have spin \bun4/3.
The level-$k$ $\CA_2$-parafermion currents $\ps_\a(z)$  generate the algebra
\refs{\rNY,\rG}
\eqn\pfope{\eqalign{
&\ps_\a(z)\ps_\b(w)\sim\frac{K_{\a,\b}}{(z-w)^{1+\a\cdot\b/k}}
     \[ \ps_{\a+\b}(w)+\ha (z-w)\der\ps_{\a+\b}(w)+\cdots\]\, ,\cr
\apar
&\ps_\a(z)\ps_{-\a}(w)\sim\inv{(z-w)^{2-\a^2/k}}\[I+
                     \frac{k+3}{3k}(z-w)^2\, T_{\rm para}(w)+\cdots\]\,,\cr}}
where $K_{\a,\b}=K_{\b,\a}$,
$K_{\a_1,\a_2}=K_{\a_1,-\t}=K_{\a_2,-\t}=1/\sqrt{k}$ and otherwise zero.
It is  then straightforward to check that\foot{
For $k=1$, this realization coincides with that considered in \rGS.}
\eqn\ZFreal{\eqalign{
&\ps(z)=\inv{\sqrt{3}}\sum_{\a=\a_1,\a_2,-\t}G^\a(z)\, ,\cr
\noalign{\vskip 2mm}
&\ps^*(z)=\inv{\sqrt{3}}\sum_{\a=\a_1,\a_2,-\t}G^{-\a}(z)\, ,\cr
\noalign{\vskip 2mm}
&T(z)=T_{\rm para}(z)-\ha(\der\vp(z))^2\, ,\cr}}
satisfy \ZFalgebra\ with
\eqn\ZFcenter{c=\frac{8k}{k+3},\qquad \la=\frac{2}{\sqrt{3k}}\, .}
The associativity is clear by construction.
This construction is rather reminiscent of making an $N=1$ superconformal
 generator out of  two $N=2$ superconformal generators.

\listrefs

\bye\bye